\documentclass[preprint]{aastex}

\begin{document}

\date{To be published in the proceedings of ``The Cosmic Microwave Background and its Polarization'', New Astronomy Reviews (eds. S. Hanany and K. A.Olive)}

\title{{\bf EXPERIENCES WITH AIRBORNE AND GROUND-BASED POLARIMETRY}}

\author{Roger Hildebrand, Chicago}

\begin{abstract}
Polarimetry in the far-infrared and submillimeter has been valuable in
tracing magnetic fields in bright Galactic Clouds. We discuss the techniques
we have developed and trends we have found in polarization $vs$ column
density, and $vs$ wavelength. The polarization spectrum has proven to be
more interesting than had been anticipated. It is potentially valuable in
explaining grain alignment and in characterizing dust species. One can
expect the infrared cirrus to be a much simpler environment than the
molecular clouds that have heretofore been explored at multiple wavelengths.
Although cirrus observations must deal with low signals and wide extent, it
is becoming possible to measure accurately the spectral energy distribution
from mid-infrared to microwave frequencies and it should soon become
feasible to determine the polarization spectrum over the same range.
\end{abstract}

\section{INTRODUCTION}

Our experiences with airborne and ground-based polarimetry have been
primarily with\ observations of bright Galactic clouds. It is possible that
some of those experiences will be relevant to CMB polarimetry. What is
certain is that instruments and observations for CMB polarimetry will be
valuable for those of us investigating magnetic fields and dust in the
Galaxy. What we can say is how we minimize the effects of atmospheric
fluctuations and what variations in polarization we observe with column
density and wavelength. Almost everything in the first part of this
presntation you can find in a {\it Primer on Far-Infrared Polarimetry}
(Hildebrand et al. 2000).

The second part of the presentation will be about requirements for more
powerful instruments in the future: about resolution, area coverage,
sensitivity, and observing techniques for more difficult goals such as
determining the nature of the IR cirrus.

\section{TECHNIQUES AND TRENDS}

\subsection{Atmospheric Fluctuations}

We begin with the problem of fluctuations in atmospheric transmission and
emission. What the {\it Primer} does not say is that our first attempt to
do far-infrared polarimetry was a failure. What we did was simply to rotate
an analyzer in front of a photometer. My advice to anyone who is serious
about polarimetry is: Don't follow that bad example. Don't build a
photometer, put a widget in front, and expect to get a first class
polarimeter. If you are observing from the ground or an aircraft, your
efforts will go largely toward coping with atmospheric fluctuations. Some
very talented observers have developed tricks for reducing the effects of
those fluctuations (e.g. Jeness, Lightfoot, \& Holland 1998), and I hasten
to acknowledge that the SCUBA polarimeter at the JCMT which does follow that
bad example is the most productive polarimeter there is at 850 $\mu m$. It
just isn't as nearly as good a polarimeter as it could be.

If you want a first class instrument to operate within the earth's
armosphere, follow instead our next example. Build a real polarimeter that
detects two components of polarization simultaneously. That can be done the
way DASI does it or the way Planck will do it. The way we do it at far-IR
and submillimeter wavelengths is to put inside the cryostat a half-wave
plate followed by a polarizing beam splitter and to provide detector arrays
for the beams reflected and transmitted by the beam splitter. The principle
of the apparatus is shown in Figure 1. 

\begin{figure}
\epsscale{0.5}
\plotone{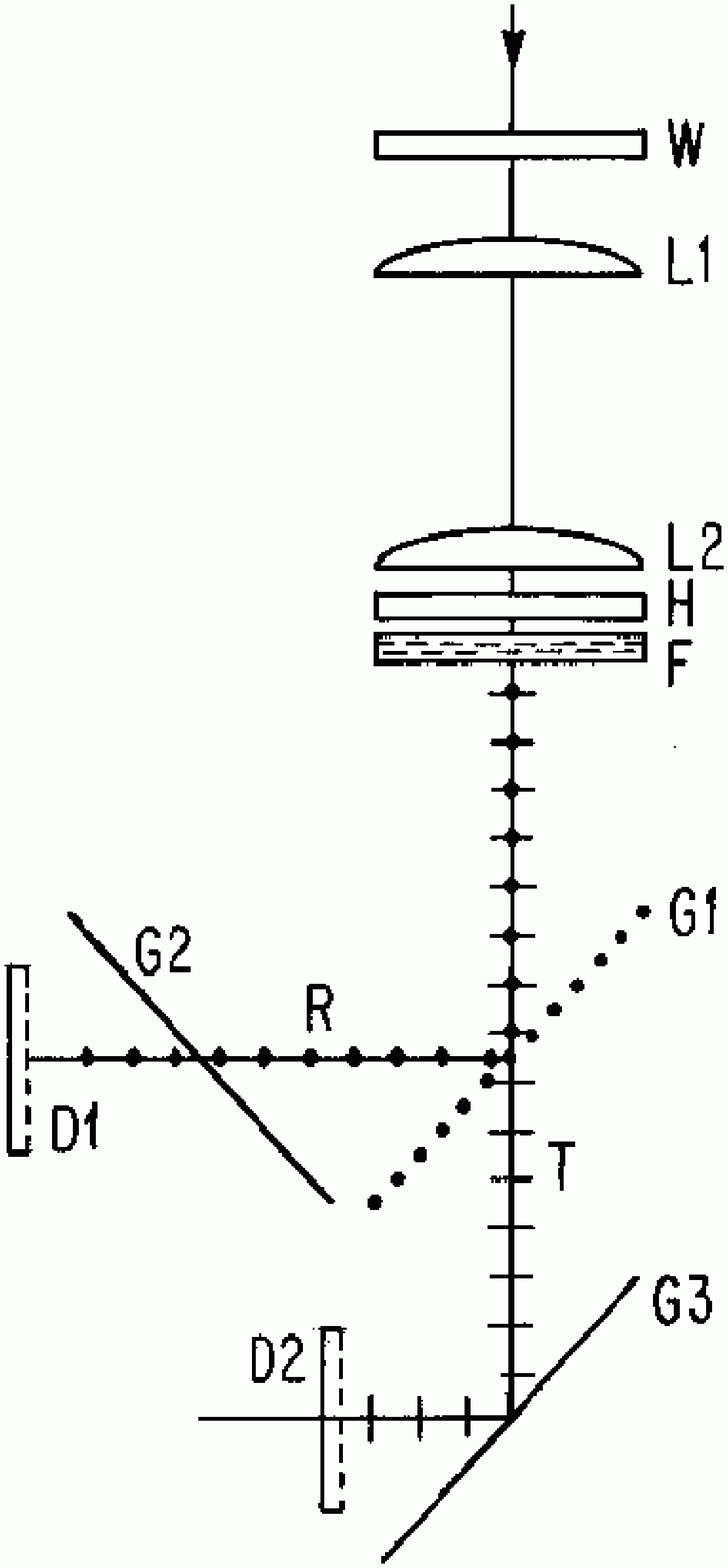}
\epsscale{1}
\caption{Schematic diagram of optics for a polarimeter. Radiation enters the
cryostat through a vacuum window, W, passes through a lens, L1, at an image
of the sky and through a second lens, L2, at an image of the primary. L2\
reimages the sky onto the detector arrays, D1 and D2. A spectral filter, F,
defines the passband. A wire grid, G1, reflects component R\ parallel to the
grid wires and transmits component T, perpendicular to the wires. Additional
grids, G2 and G3, ensure that each beam is reflected once and transmitted
once before reaching the detectors. One measures the quantity $(R-T)/(R+T)$
as a fuction of the angle of rotation of the half-wave plate, H.}
\end{figure}

If you chop between points on and off of an object and then, more slowly,
nod the telescope so as to put that object first in one beam and then in the
other, what you would like to see for each component is a square wave. What
you actually see for a faint source is a square wave greatly distorted by
fluctuations in the transmission and emission of the atmosphere. Figure 2
shows a strip chart record for $\thicksim $5 minutes of observations of IRC+10216. The flux from this source is ten times the photon-noise-limited NEFD and comparable to the atmospheric fluctuations on $\thicksim $1 minute time scales. Since the fluctuations affect both components simultaneously, they are reduced by more than an order of magnitude when you take the difference of the signals
over the sum. The correlation in the noise between the two components is
evident. Notice the correlated drifts in the baseline as well as the spikes
appearing in each nod pair. The drifts are much larger over a period of an
hour. Keep in mind that the radiation reaching the instrument comes
primarily from the atmosphere and telescope, not from the object of interest.

Fluctuations in transmission affect the two components by equal factors.
Fluctuations in emission affect the components by equal increments and hence
are not completely removed when taking the difference in the two components
over the sum. The analysis system for faint sources observed over long
periods must take that into account. 

\begin{figure}
\plotone{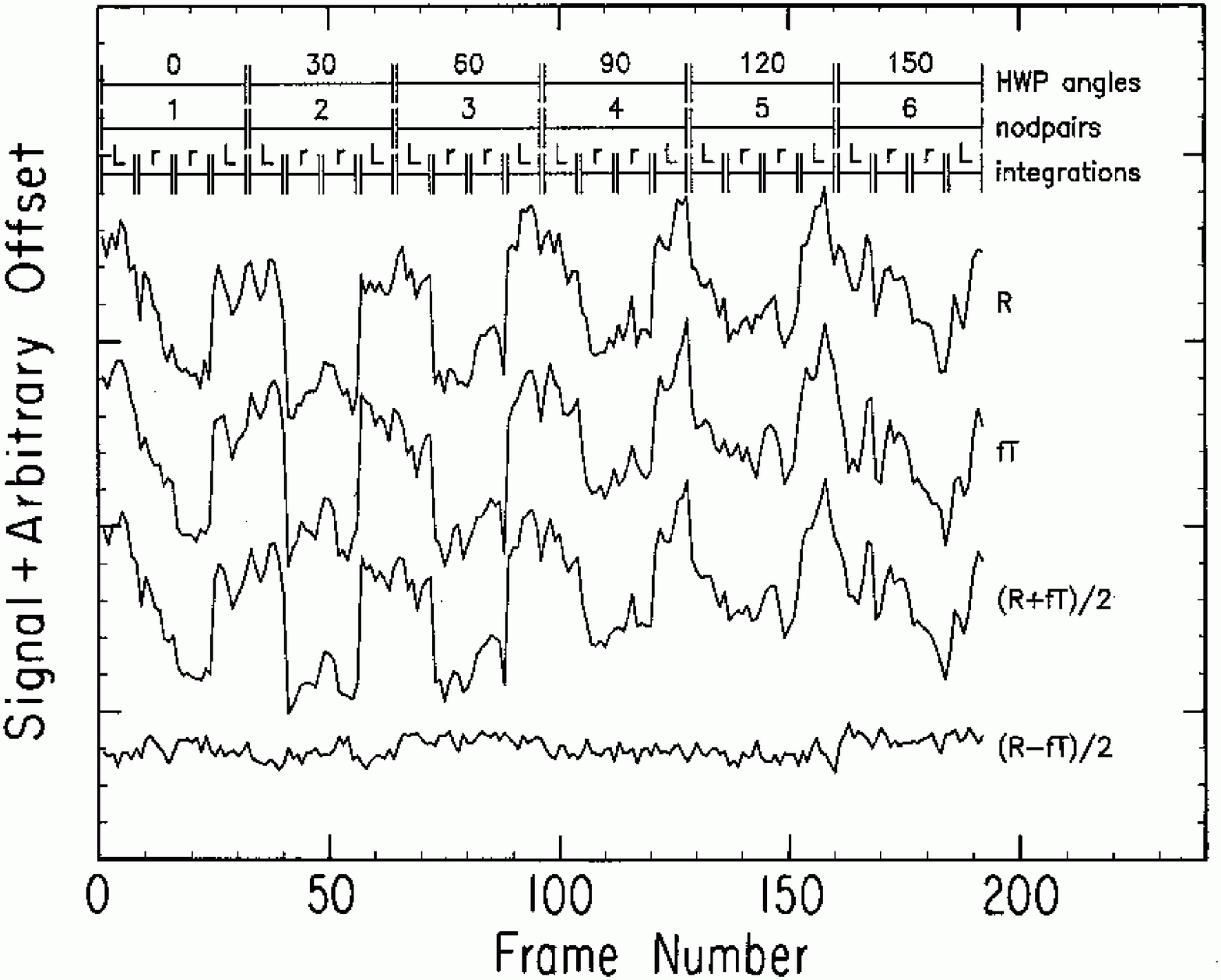}
\caption{Strip chart record (Dowell et al. 1998) of frames accumulated during a
single file (6 steps of the half-wave plate) during an observation of the
source IRC+10216 (F$_{\protect\nu }=30$ Jy). The $T$-frames have been
multiplied by a normalization factor, $f$, to bring them to the same scale
as the$R$-frames. The correlated sky noise is removed by taking the
difference, $R-fT$.}
\end{figure}

If the source is polarized the amplitude of the square wave should be
modulated as you turn the halfwave plate. You can try analyzing the results using only one component at a time. For a faint unpolarized source, the signals
for the individual components vary erratically as shown in the top two
panels of Figure 3.\ If instead you do it properly, taking the difference
over the sum, a sinusoidal variation\ giving the amplitude and phase of the
polarization appears as shown in the bottom panel.

\begin{figure}
\epsscale{0.5}
\plotone{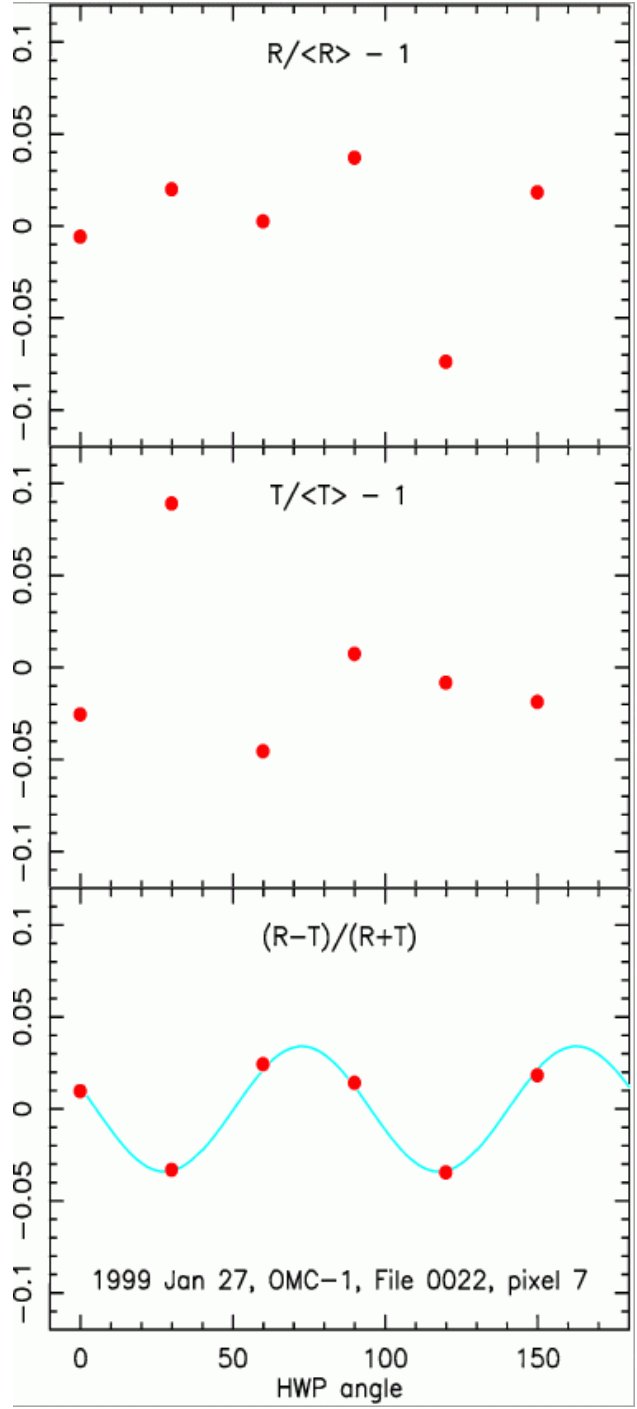}
\epsscale{1}
\caption{Effect of atmospheric variations on polarization
signals. The top two panels show the results obtained by analyzing
separately the data for the two orthogonal compnents of polarization for a
single rotation of the half-wave plate through 180$^{\circ }$ in 30$^{\circ }
$ steps. As is evident, no satisfactory fits are possible. The bottom panel
shows the fit obtained when the data from the two components are combined to
form $(R-T)_{\protect\theta }/(R+T)_{\protect\theta }$.}
\end{figure}

If your apparatus is in a balloon at 120,000 ft., it will float with the
atmosphere and the problem of rapid fluctuations will be much less severe,
but long-term drifts will still be a problem.

\subsection{Degrees of Polarization vs Optical Depth}

We have published an archive of results from airborne polarimetry at 60 $\mu
m$ and 100 $\mu m$ (Dotson et al. 2000). Examples of our ground-based
observations from the Caltech Submillimeter Observatory (CSO) at 350 $\mu m$
are shown in Figure 4. These maps were made with 32-pixel dector arrays
where the performance of the arrays was far below the current state of the
art and were made at a wavelength where the degree of polarization is near a
minimum (Fig. 6A). Nevertheless significant detections were found at nearly
every point. The flux density contours are derived from the sums of the
reflected, $R$, and transmitted, $T$, signals.

\begin{figure}
\plotone{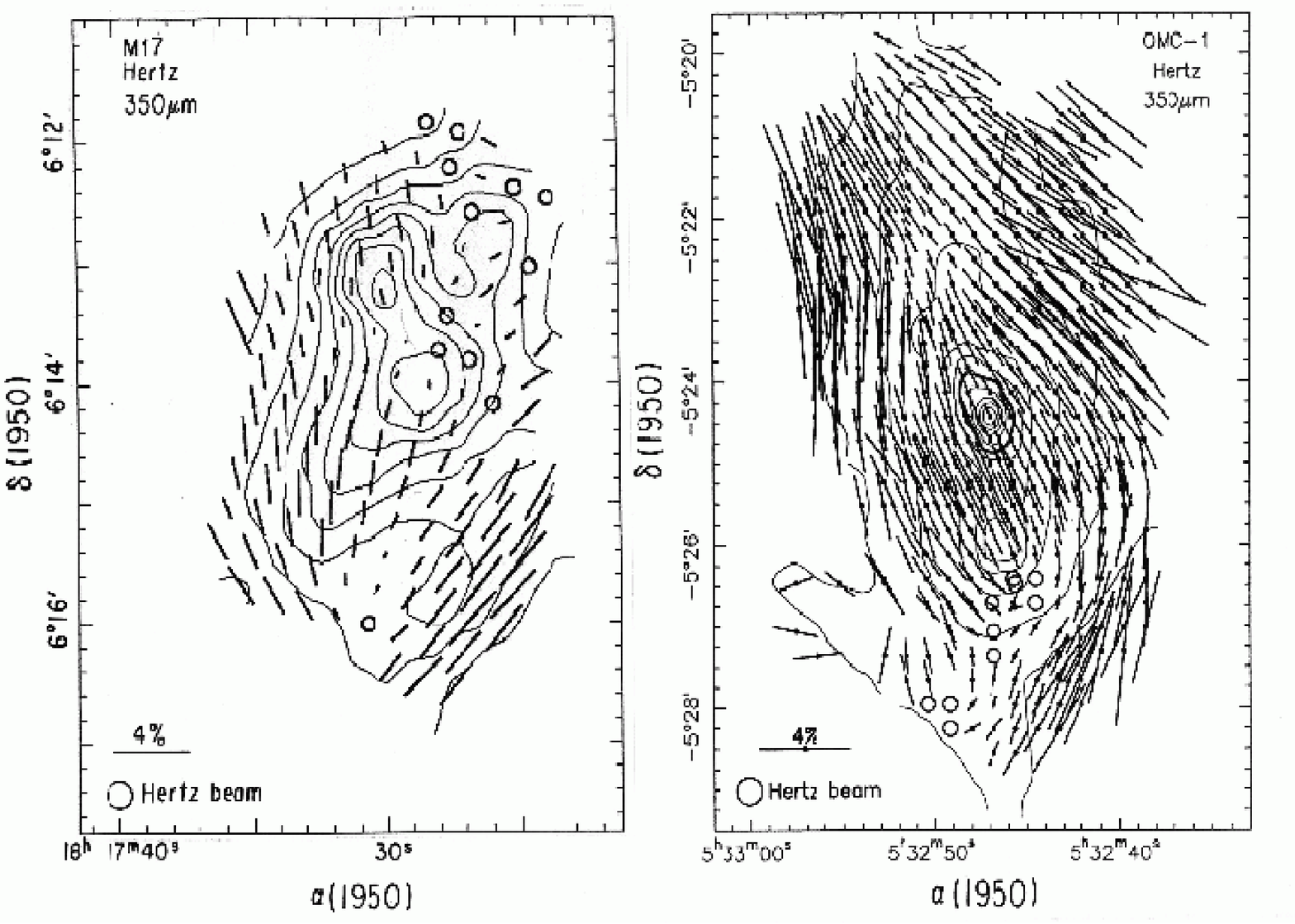}
\caption{Polarization maps of M17 and
OMC-1 (E-vectors). All vectors are for $\geq 3\protect\sigma $ measurements.
Open circles denote points at which measured values of $P$ were below $3%
\protect\sigma $ and at which $P+2\protect\sigma <1\%.$ }
\end{figure}

Notice that the degree of polarization, denoted by the length of the vectors
in Figure 4, tends to increase toward the lower contours. Figure 5 shows the
effect for several molecular clouds. One expects a drop in the degree of
polarization at high optical depths $(\tau \gtrsim 1)$ due to opacity, but
the polarization starts to drop at lower optical depths due to turbulence in
the aligning field, as described by Terry Jones at this workshop (see also
Jones, Klebe, \& Dicke 1992) and, in addition, may increase as $\tau
\rightarrow 0$ if, as we suspect, the conditions for grain alignment are
more favorable at low column densities.

\begin{figure}
\plotone{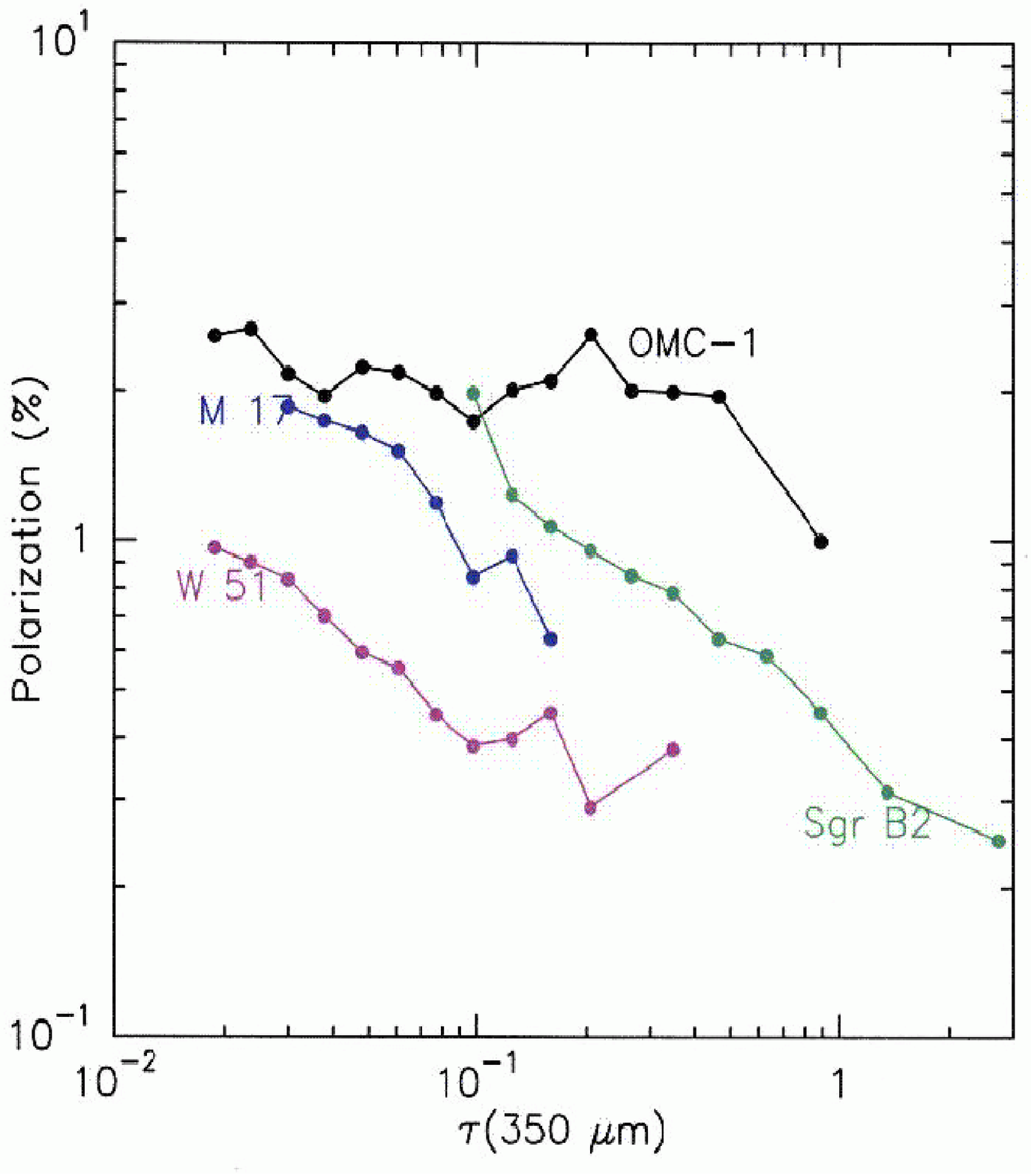}
\caption{Degree of polarization $vs$ optical depth for four clouds.}
\end{figure}

\begin{figure}[h]
\plotone{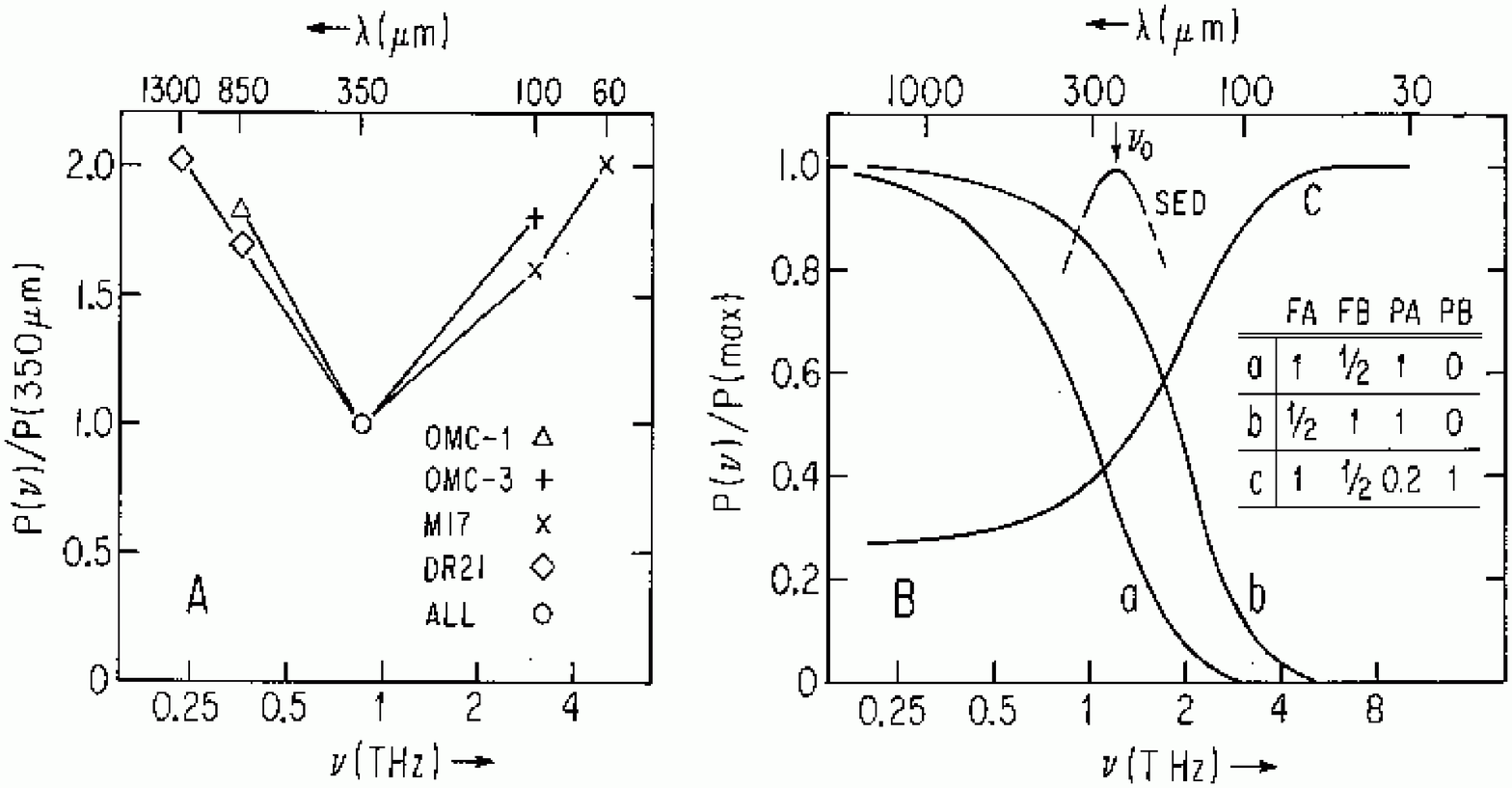}
\caption{Polarization Spectra: (A) as observed
in molecular clouds (Vaillancourt 2002), and (B) as predicted for diffuse
Galactic emission: three possibilities are shown. FA, FB = relative peak
flux densities and PA, PB = relative polarization efficiencies for
components A, B. $\protect\nu _{0}=$ observed peak frequency of the spectral
energy distribution.}
\end{figure}

\subsection{Wavelength-Dependence}

In molecular clouds we find a substantial dependence of polarization on
wavelength (Hildebrand et al. 1999; Hildebrand et al. 2000; Hildebrand 2002)
in the range 60$\mu m$ to 1300 $\mu m$ (Fig. 6A). We interpret
wavelength-dependence as evidence that different populations of dust are
sampled at different wavelengths, either because domains within a cloud are
at different temperatures and have environments more or less favorable to
grain alignment, or because grain species within the clouds have different
polarizing powers and emissivities.

Molecular clouds are inherently complex. They contain embedded stars, clumps
opaque to stellar radiation, surfaces exposed to the interstellar radiation
field, and often adjacent HII regions. Their polarization spectra reflect
that complex structure. One can expect the polarization spectra of the
infrared cirrus to be much simpler if in those clouds all grains are in the
same environment. If we adopt the dust temperatures and spectral indices in
the two-component model of high latitude cirrus by Finkbeiner, Davis, and
Schlegel (1999), then we can use the polarization spectrum and spectral
energy distribution to determine the relative abundances, polarizing powers,
and spectral indexes of the components. Figure 6B gives examples.

\section{FUTURE POLARIMETERS}

\subsection{Multiple Passbands}

The $\lambda $-dependence of $P$ shown in Figure 6A is a recent discovery
(Hildebrand et al. 1999; 2000; Vaillancourt 2002) based on results from
three different instruments, one of which had to be reconfigured to reach 60 $%
\mu m$ (Dowell 1997). To characterize the dust emission in the diffuse
Galactic foreground one needs spectral energy distributions and
polarization spectra over a wider range of wavelengths than is likely to be
spanned by single instruments. The analysis calls for development of
additional polarimeters with multiple passbands over the range $\thicksim $%
50 $\mu m$ to $\thicksim $3 $mm$. Data from the WMAP archive (Bennett et al.
2003) show the value of multiple passbands at microwave frequencies (Lagache
(2003). An important result of microwave observations is that there are
large variations in the spectrum from point to point (Finkbeiner, et al.
2002; de Oliveira-Costa et al. 2002).

\subsection{Sensitivity: Polarimetry of IR-Cirrus}

With Hale, a new polarimeter on SOFIA, we expect a sensitivity at 100$\mu m$ of $\thicksim 1Jy/
\sqrt{Hz}$ at 100$\mu m$ and should thus be well within range not only for photometry but also for polarimetry assuming $\mid P\mid $ $=$ a few per cent. High latitude cirrus clouds are often as bright as a few times 10 $MJy/Sr$, a difficult but probably manageable level. A
complication will be that one cannot put the reference beams in cirrus-free regions. It will be necessary to record fluctuations in polarization $vs$ angle as in microwave background observations. One must be content with measuring power spectra in the degree and angle of polarization taking into account the large-scale magnetic field along the Galactic arms. Fosalba et al. (2002) have measured the angular power spectrum, $C_{l}$, of starlight
polarization for the Galactic plane ($\mid b\mid <10^{\circ })$. They find $%
C_{l}\varpropto l^{-1.5}$ where $l\thickapprox 180^{\circ }/\theta $. High
latitude cirrus tend to be closer and hence the power spectra should have
reduced high frequenciy components.

\subsection{Array Size, Beam Size}

No detector arrays for the first round instruments on SOFIA come close to
filling the useful focal plane with diffraction-limited and
background-limited detectors. But multiplexed read-out systems are reducing
the volume required for cold electronics. It may be feasible to install two
arrays, each with $>$ 1000 TES pixels into the same volume now
occupied by one $12\times 32$ (384-pixel) array of semi-conductor
bolometers. No large multiplexed arrays
have yet been proven, but expert groups are working at the problem and one
can anticipate success in the next year or two.

Our polarimeters, Mark I, Stokes, and Hertz, (Dragovan 1986; Platt et al.
1994; Schleuning et al. 1997; Dowell et al. 1998) have not had
diffraction-limited beams. Hertz at the CSO, has 20 arcsec beams but with a
proposed upgrade it will have 8 arcsec beams. Hale on SOFIA will have 20
arcsec beams at 200 $\mu m$, 10 arcsec beams at 100 $\mu m$, and 5 arcsec
beams at 50 $\mu m$. These may be smaller than are interesting for cosmic
background observations but not too small for compact objects such as Sgr A*
and protostellar cores.

\section{\protect\bigskip Summary}

Our experiences and recommendations relevant to this workshop can be
summarized as follows:

1. Minimize effects of atmospheric fluctuations: Observe two components
simultaneously. (Figs. 2 \& 3)

2. $\mid P\mid $ increases as $\tau $ decreases (Figs. 3 \& 4).

3. $\mid P\mid $ $=P(\lambda )$ (Fig 6A). We expect variations in IR cirrus
unlike those in molecular clouds (Fig 6B)

4. Provide multiple passbands - {\it Calibrate}{\bf \ }both flux
density and polarization measurements

5. To the extent possible, fill the focal plane with diffraction-limited and
background-limited pixels.

6. For extended objects (e.g. CMB, IR cirrus), measure angular power spectra
in $\mid P\mid $ and in the polarization angle.

7. To characterize the polarized dust emission of the cosmic foreground,
measure the spectral energy distribution and polarization spectra of IR
cirrus over at least the range $\thicksim $60$\mu m$ to a few mm.

\subsection{Acknowledgements}

I am grateful to my student, Larry Kirby, for help in preparing this
manuscript. I thank my former students Jackie Davidson, Jessie Dotson,
Darren Dowell, Mark Dragovan, Giles Novak, David Schleuning, and John
Vaillancourt for their contributions to the airborne and ground-based work
summarized here. I thank Darren Dowell for Figures 2, 3, and 5. Figures 1
and 2 are reproduced with permission from Hildebrand et al. 2000, PASP, 112,
1215. This work has been supported by NSF Grants AST-9987441 and AST-0204886.

\section{REFERENCES}

\begin{description}
\item Bennett, C. L., Hill, R. S., Hinshaw, G., Nolta, M. R., Odegard, N.,
Page, L., Spergel, D. N., Weiland, J. L., Wright, E. L., Halpern, M.,
Jarosik, N., Kogut, A., Limon, M., Meyer, S. S., Tucker, G. S., \& Wollack,
E., 2003, Submitted to ApJ

\item de Oliveira-Costa A., et al. 2002, ApJ, 567, 363

\item Dotson, J. L., Dowell, C. D., Schleuning, D. A., \& R. H.
Hildebrand. 2000, ApJS, 128, 335 - 370

\item Dowell, C. D. 1997 ApJ, 487, 237

\item Dowell, C. D., Hildebrand, R. H., Schleuning, D. A., Vaillancourt, J.
E., Dotson, J. L., Novak, G., Renbarger, T., \& Houde, M. 1998, ApJ, 504, 588

\item Draine, B. T. \& Lazarian, A. 1998 in{\it\ Microwave Foregrounds }%
\ Ed's, Angelica de Oliveira-Costa and Max Tegmark. Astronomical Society of
the Pacific Conf. Series Vol 181

\item Draine, B. T., and Lazarian, A. 1998a, ApJ, 494, L19

\item Draine, B. T., and Lazarian, A. 1998b, ApJ, 508, 157

\item Dragovan, M. 1986, ApJ, 308, 270

\item Finkbeiner, D. P., Davis, M., \& Schlegel, D. J. 1999, ApJ, 524,867

\item Finkbeiner, D. P.,Schlegel, D. J., Frank, C., \& Heiles, C. 2002, ApJ,
566, 898

\item Fosalba, P., Lazarien, A., Prunet, S., \& Tauber, J. A. 2002, ApJ,
564, 762

\item Hildebrand, R. H., Dotson, J. L., Dowell, C. D., Schleuning D. A., \&
Vaillancourt 1999, ApJ, 516, 834

\item Hildebrand, R. H., Davidson, J. A., Dotson, J. L., Dowell, C. D.,
Novak, G., and Vaillancourt. J. E. 2000, PASP, 112, 1215. See Errata 2000,
PASP, 112, 1620

\item Hildebrand, R. H. In ``Astrophysical Spectropolarimetry'' Ed: J.
Trujillo-Bueno, F., Moreno-Insertis, and F. Sanches. Cambridge University
Press (2002). pp 265 -- 302.

\item Jenness, T., Lightfoot, J. F., \& Holland, W. S. 1998, Proc. SPIE,
3357, 548

\item Jones, T. J., Klebe, D. K., \& Dickey 1992, ApJ, 389, 602

\item Lagache, G. 2003. A\&A, In press. arXiv:astro-ph/0303335 v1 \ \ 14 Mar
2003

\item Platt, S. R., Dotson, J. L., Dowell, C. D., Hildebrand, R. H.,
Schleuning, D. A. \& Novak, G. 1994, in Proc. Airborne Astronomy Symposium
on the Galactic Ecosystem: From Gas to Stars to Dust. ed. M. R. Haas, J. A.
Davidson, and E. F. Erickson. PASP

\item Schleuning, D. A., Dowell, C. D., Hildebrand, R. H., Platt, S. R., \&
Novak, G. 1997, PASP 109, 307

\item Vaillancourt, J. E. 2002, ApJS, 142, 53
\end{description}

\end{document}